# Interpretable self-driving sputter epitaxy: from black-box optimization to human-usable growth rules

Yuki K. Wakabayashi,[1,*] Yui Ogawa,[1,*] Franz Benedict Romero,[1] Takuma Otsuka,[2] and Yoshitaka Taniyasu[1]

[1]*Basic Research Laboratories, NTT, Inc., Atsugi, Kanagawa 243-0198, Japan*
[2]*Communication Science Laboratories, NTT, Inc., Soraku-gun, Kyoto 619-0237, Japan*

[*]Correspondence: yuuki.wakabayashi@ntt.com, yui.ogawa@ntt.com

**Abstract**

Self-driving laboratories have emerged as powerful tools for navigating high-dimensional process spaces, yet systems remain black-box optimizers that yield limited transferable process understanding. Here, we demonstrate an interpretable self-driving laboratory framework that transforms autonomous optimization into human-usable growth rules. As a stringent benchmark, we apply this framework to RF magnetron sputtering, addressing a long-standing challenge of achieving high-quality $β$-Ga$_2$O$_3$ heteroepitaxy and single-crystalline $β$-Ga$_2$O$_3$ homoepitaxy via sputtering. By combining Bayesian optimization with automated optical evaluation of the Urbach energy as a metric of sub-bandgap disorder, the self-driving system efficiently identifies heteroepitaxial growth conditions yielding a minimum Urbach energy of 182 meV, the lowest value for sputtered $β$-Ga$_2$O$_3$ films. Importantly, the optimized growth window is transferable, realizing single-crystalline $β$-Ga$_2$O$_3$ homoepitaxy without further optimization, corroborated by scanning transmission electron microscopy. To convert the closed-loop dataset into interpretable growth rules, we train a random forest surrogate and distill it into response curves and quantified pairwise interactions across the four-dimensional growth-parameter space. This analysis identifies substrate temperature as the primary control knob, with RF power and gas flows acting largely additively and only a modest temperature-oxygen coupling delineating the narrow window for high-quality growth, establishing a general route from autonomous experimentation to transferable growth rules.

**Introduction**

Self-driving laboratories (SDLs) are rapidly transforming materials research by closing the loop between automated experimentation[1,2] and machine-learning (ML) decision making[3-5], with Bayesian optimization (BO)[6-11] emerging as a particularly efficient engine for navigating high-dimensional process spaces[12-17]. Yet, despite their success at identifying optimal recipes with fewer trials, most SDL implementations still function as black-box optimizers[18-20]: they output optimal conditions but do not reveal which process parameters fundamentally govern performance, how each knob reshapes the response, or whether the system is controlled by simple additive trends versus complex process parameter interactions. This interpretability gap directly undermines reproducibility, limits transferability across tools, substrates, and material systems, and slows the conversion of autonomous optimization into robust, physically meaningful process knowledge. Addressing this gap requires self-driving frameworks that go beyond recipe discovery to expose dominant controls and parameter interactions in a human-usable form. Such interpretable SDLs are essential to bridge autonomous optimization and transferable process engineering.

A stringent benchmark for an interpretable SDL is the sputter epitaxy of $\beta$-$Ga_2O_3$ films, a promising ultra-wide-bandgap semiconductor for next-generation power electronics and solar-blind deep-ultraviolet photodetectors, where device performance is highly sensitive to crystalline quality and optical disorder[21-27]. High-quality epitaxial $\beta$-$Ga_2O_3$ has been achieved by established epitaxy techniques such as chemical vapor deposition (CVD)[28-30] and molecular beam epitaxy (MBE)[31-34]; however, achieving comparably high-quality epitaxy by sputtering, which is an industry-standard, low-cost route for large-area deposition, remains elusive[35-37]. The energetic growth environment and limited adatom mobility, together with narrow phase-stability margins in complex oxides, make epitaxial control challenging. As a result, sputtered $Ga_2O_3$ films are often amorphous, polycrystalline, or mixed-phase (e.g., $\varepsilon$- and $\kappa$-$Ga_2O_3$)[35, 38-40]. These characteristics make sputtered $\beta$-$Ga_2O_3$ an ideal and stringent testbed for SDLs: the challenge is not only to achieve high-quality epitaxy by sputtering, but also to go beyond black-box optimization and extract interpretable, actionable, and transferable growth rules from the closed-loop data.

Here, we address this challenge by realizing an interpretable SDL for the sputter epitaxy of *β*-Ga$_2$O$_3$ films. We construct a self-driving sputtering platform that couples automated RF magnetron sputtering with automated optical analysis and Bayesian optimization of the Urbach energy ($E_U$)[41,42], a metric of band-tail states and sub-bandgap optical disorder. Using this framework, we rapidly identify a high-quality heteroepitaxial growth window for *β*-Ga$_2$O$_3$ on C-plane Al$_2$O$_3$ (0001), achieving a minimum $E_U$ of 182 meV by the 56th run (66 total closed-loop runs). The resulting film exhibits high crystalline quality, evidenced by a narrow *β*-Ga$_2$O$_3$ ($\bar{2}$01) rocking-curve full width at half maximum (FWHM) of 1.78°, as well as high optical quality, reflected in the low Urbach energy, together with a smooth surface with 1.15–1.27 nm roughness. Importantly, the BO-identified growth window is directly transferable to *β*-Ga$_2$O$_3$ substrates, enabling single-crystalline homoepitaxy without additional optimization. This cross-substrate transferability, achieved without retuning, indicates that the self-driving process captures intrinsic growth rules rather than substrate-specific recipes. Notably, to our knowledge, single-crystalline *β*-Ga$_2$O$_3$ homoepitaxy by sputtering has not been reported previously[35]. Beyond recipe discovery, we introduce an interpretable machine-learning analysis workflow that interrogates a tree-based surrogate model to distill closed-loop data into human-usable growth rules. This workflow identifies substrate temperature as the dominant control parameter, with RF power and gas flows providing largely additive secondary tuning and only a modest temperature-oxygen coupling that defines the narrow window for high-quality growth. This structure suggests a human-usable optimization strategy: sequential one-dimensional tuning followed by a final two-dimensional refinement in the ($T$, O$_2$) plane. Together, these results establish an interpretable SDL framework that transforms autonomous experimentation into transferable process understanding beyond black-box optimization.

**Results**

**Self-driving sputter growth with automated optical analysis**

We establish an interpretable self-driving sputtering workflow in which a physically meaningful disorder metric guides closed-loop exploration of a high-dimensional growth-parameter space. Figure 1 summarizes the workflow in this study. *β*-Ga$_2$O$_3$ films with a thickness of 55 nm were grown on a 2-inch C-plane Al$_2$O$_3$ wafer by RF magnetron sputtering while varying four growth parameters: substrate temperature *T*,

RF power $P_{RF}$, and Ar and $O_2$ flow rates ($F_{Ar}$, $F_{O_2}$) (Fig. 1a) (see Methods section "Self-driving sputtering system" and Supplementary Fig. 1). After each deposition, the transmittance spectrum of the film was measured (Fig. 1b) and automatically analyzed to extract the Urbach energy $E_U$ (Fig. 1c) (see Methods section "Automated optical analysis"), which serves as a sensitive figure of merit for band-tail states and optical disorder in ultra-wide-bandgap semiconductors[41,43]. The extracted value was fed back to a Gaussian-process (GP)-based BO algorithm tailored to thin-film growth under occasional experimental failures (see Methods section "Bayesian optimization with experimental failures and adaptive GP prior mean")[44,45] which proposed the next set of growth parameters and automatically initiated the subsequent growth run on the next wafer in the stock, thereby forming a semi-autonomous closed loop. In the present implementation, each loop consists of (i) manual loading of up to five 2-inch wafers into the load-lock (wafer stock), (ii) fully automated, vacuum-integrated transfer between the load-lock and the sputter growth chamber by a built-in robot arm, sputter growth according to the BO-suggested parameters, and automatic return to the load-lock, (iii) manual transfer of the grown film from the load-lock to the optical setup, and (iv) fully automated optical measurement and spectral analysis, followed by BO model update and next-point selection that automatically triggers the next sputter growth. Because wafer handling inside the sputter system is already robotized, the remaining manual transfer step in (iii) could also be replaced by robotic handlers, so that the current semi-autonomous workflow can be straightforwardly extended to a fully autonomous self-driving sputter growth platform. Finally, an interpretable machine-learning analysis was applied to the BO dataset to extract a simple, human-readable rule for $E_U$ (Fig. 1d).

Across successive self-driving runs, the system explores and refines growth conditions, resulting in a steady reduction in $E_U$. Figure 2a shows the evolution of the $E_U$ during the 66 runs of the self-driving sputter experiments. The first seven runs used a pragmatic baseline set of parameters (fixed $P_{RF}$, $F_{Ar}$ and $F_{O_2}$) and swept the substrate temperature over 30–700 °C to seed BO. Starting from these initial data, the BO algorithm proposed the growth parameters for the subsequent 59 runs at each iteration. As the growth run number increases, the grown films have a broad range of $E_U$ values, reflecting exploratory proposals by BO, whereas the lowest $E_U$ values at that time decrease and eventually reach 182 meV at ($T$, $P_{RF}$, $F_{Ar}$, $F_{O_2}$) = (507°C, 119 W, 26 sccm, 1 sccm). This is the lowest $E_U$ reported for sputtered $β$-Ga$_2$O$_3$ films, improving on the previous best value of 280 meV[47], and it is also lower than a representative metal-organic chemical vapor deposition (MOCVD)-grown report (220 meV)[46] (Table 1)[31,46-65], highlighting that self-driving optimization can push sputtered $β$-Ga$_2$O$_3$ into a markedly improved optical-quality regime. A replication growth performed at the identified optimum conditions yielded an $E_U$ of 184 meV, consistent with the best run (182 meV), indicating high reproducibility of the optimum.

Mapping the measured $E_U$ across the multidimensional growth-parameter space reveals structured low-disorder regions interspersed with exploratory sampling. Figures 2b-2g map experimental $E_U$ onto the $P_{RF}$-$F_{Ar}$-$F_{O_2}$ space across different temperature ranges. We observe dense clusters of low-$E_U$ films around sweet spots in the 400–500°C and 500–600°C regions, while sparsely distributed high-$E_U$ points in all temperature regions appear as BO explores in the four-dimensional growth space. This feature clearly reflects the characteristic balance between exploiting known favorable conditions and exploring new ones in the BO algorithm. The quantitative dependence of $E_U$ on the growth parameters in this high-dimensional space, as well as correlations between the parameters, can be distilled into an intuitive, human-interpretable picture by post hoc interpretable ML analysis of the same dataset, as described later.

**Crystallographic and optical properties of optimized films**

Optimizing $E_U$ is accompanied by a systematic sharpening of the absorption edge and suppression of sub-bandgap absorption in sputtered $β$-Ga$_2$O$_3$ films. Figure 3a shows the optical absorption behavior of $β$-Ga$_2$O$_3$ films on a 2-inch C-plane Al$_2$O$_3$ wafer spanning a wide range of $E_U$. The optimized film ($E_U$ = 182 meV) exhibits the steep absorption edge with a band gap energy ($E_g$) of 5.04 eV, and a strong above-bandgap absorption. Notably, the absorption coefficient at 5.5 eV reaches $α = 1.5 × 10^5$ cm$^{-1}$, which is higher than a representative value reported for MOCVD-grown $β$-Ga$_2$O$_3$ films ($α ≈ 8.4 × 10^4$ cm$^{-1}$ at 5.5 eV)[66], highlighting the high optical quality of the optimized film. In contrast, films with larger $E_U$ show a pronounced sub-bandgap (in-gap) absorption tail. When the spectra are normalized (inset), increasing $E_U$ systematically enhances the in-gap absorption and shifts the nominal $E_g$ to lower photon energies (e.g., 4.86 eV for $E_U$ = 492 meV). These results confirm that using $E_U$ as the objective in BO directly leads to a substantial improvement in optical quality.

Consistent with the optical improvement at lower $E_U$, the $E_U$-optimized films also exhibit enhanced crystallographic quality. The crystalline quality of the BO-optimized $β$-Ga$_2$O$_3$ film was examined by X-ray diffraction (XRD) and compared with that of a non-optimized film ($E_U$ = 304 meV) (Fig. 3b). For the optimized $β$-Ga$_2$O$_3$ film, intense and well-defined peaks from the ($\bar{2}$01)-oriented $β$-Ga$_2$O$_3$ phase are observed[60], while no $κ$-Ga$_2$O$_3$ or other secondary phases are detected within the resolution of the measurement, confirming single-phase $β$-Ga$_2$O$_3$ growth. In contrast, for the non-optimized film, the intensities of the $β$-Ga$_2$O$_3$ ($\bar{4}$02) reflections are about one order of magnitude weaker and barely emerge from the background, indicating poor crystalline quality. This clear enhancement of the $β$-Ga$_2$O$_3$ peak intensities alongside the reduction in $E_U$ indicates a strong correlation between optical disorder and the crystalline quality in the sputtered films. The rocking curve of the $β$-Ga$_2$O$_3$ ($\bar{2}$01) reflection [Fig. 3c] is well described by a Lorentzian profile with a FWHM of 1.78°. For comparison, various CVD-based and MBE-based techniques report FWHM values in the range of ≈ 0.33–2.0° for $β$-Ga$_2$O$_3$ films on C-plane Al$_2$O$_3$ (Table 1). Our sputter-grown film with a FWHM of 1.78° thus lies well within the range of high-quality $β$-Ga$_2$O$_3$ grown by CVD- and MBE-based techniques, even though it is obtained by RF magnetron sputtering rather than a dedicated and complex epitaxial reactor or ultrahigh vacuum.

The optimized film exhibits well-defined heteroepitaxial registry with the C-plane Al$_2$O$_3$ substrate, as evidenced by in-plane XRD and atomic-resolution scanning transmission electron microscopy (STEM) analyses. To reveal the in-plane orientation of the $β$-Ga$_2$O$_3$ film relative to the C-plane Al$_2$O$_3$ substrate, the XRD-$φ$ scan was performed using the $β$-Ga$_2$O$_3$ ($\bar{4}$01) reflection (Fig. 3d). The $φ$ scan exhibits six peaks separated by 60°, indicating heteroepitaxial in-plane alignment on C-plane Al$_2$O$_3$. Considering the twofold symmetry of monoclinic $β$-Ga$_2$O$_3$, these six peaks correspond to three equivalent in-plane rotational domains of the ($\bar{2}$01)-oriented $β$-Ga$_2$O$_3$ film. Such multi-domain in-plane epitaxy on C-plane Al$_2$O$_3$ has also been reported for high-quality heteroepitaxial $β$-Ga$_2$O$_3$ grown by CVD-based methods, consistent with the established heteroepitaxial relationship $\beta$-Ga$_2$O$_3$ [102] ∥ Al$_2$O$_3$ ($1\bar{1}20$)[49,56]. Figure 3e shows an atomic-resolution high-angle annular dark-field (HAADF)-STEM image of one of these in-plane domains, and the detailed multi-domain in-plane epitaxial structure is further confirmed in Supplementary Information, Section II. Whereas previous RF-sputtered Ga$_2$O$_3$ films on C-plane Al$_2$O$_3$ have largely been limited to amorphous, polycrystalline, or strongly textured films[35,36,37], achieving a single-phase $\beta$-Ga$_2$O$_3$ epilayer with a low $E_U$ and narrow rocking-curve FWHM thus marks a substantial advance in sputter-based growth.

In addition to improved optical and crystallographic quality, the $E_U$-optimized $\beta$-Ga$_2$O$_3$ film also shows a flat surface, with a root mean square (RMS) roughness of 1.15–1.27 nm (Fig. 3f). This value is comparable to the lower end of the RMS roughness reported for heteroepitaxial $\beta$-Ga$_2$O$_3$ films grown by CVD- and MBE-based techniques (typically 0.21–7 nm) (Table 1). The achieved surface smoothness is expected to be beneficial for heteroepitaxial device fabrication, where reduced interface roughness can help mitigate carrier scattering and improve breakdown and reliability.

**Homoepitaxial growth on $\beta$-Ga$_2$O$_3$ substrates**

To test whether the BO-optimized window identified for heteroepitaxial growth captures intrinsic growth trends rather than substrate-specific effects, we apply the same window to $\beta$-Ga$_2$O$_3$ substrates and demonstrate single-crystal homoepitaxy by sputtering for the first time. A low-magnification cross-sectional HAADF-STEM image (Fig. 4a) reveals a laterally uniform $\beta$-Ga$_2$O$_3$ film. A higher-magnification HAADF-STEM image (Fig. 4b) shows a continuous array of well-ordered atomic columns without grain boundaries or rotational domain boundaries, indicating single-crystal homoepitaxial growth. Atomic-resolution STEM images in annular bright-field (ABF) (Fig. 4c) and HAADF (Fig. 4d) modes further resolve the $\beta$-Ga$_2$O$_3$ cation and anion sublattices with excellent ordering: in the ABF image, both Ga and O columns are visible, whereas in the HAADF image, the heavier Ga columns dominate the contrast, with the projected lattice matching the expected crystal structure of $\beta$-Ga$_2$O$_3$[67]. These observations show that the self-driving optimization performed on C-plane Al$_2$O$_3$ directly identifies a growth window that also enables single-crystal $\beta$-Ga$_2$O$_3$ homoepitaxy by sputtering without any additional optimization, establishing a new capability for sputter epitaxy.

AFM observations further confirm a smooth, pit-free surface for the homoepitaxial film: the RMS roughness is 1.15–1.27 nm (Supplementary Fig. 5), which is favorable for subsequent doping/contact processing and for minimizing interface scattering in device heterostructures. The electrical properties of the homoepitaxial $β$-$Ga_2O_3$ films were evaluated using simple two-terminal test structures. Unintentionally doped $β$-$Ga_2O_3$ layers with a thickness of 55 nm were grown under the optimized condition, and Ti (20 nm)/Au (150 nm) contacts were deposited on the film surface, followed by rapid thermal annealing at 500 °C for 1 min in $N_2$, which is a standard technique to form Ohmic contacts to $β$-$Ga_2O_3$[68,69]. However, the device resistance remained beyond the measurement range (10 GΩ or higher) over the applied voltage range up to 10 V, indicating that the optimized sputtered films are very lightly doped. This low unintentional doping level, consistent with the absence of extended defects (e.g., grain boundaries or secondary phases) in the STEM observations, is promising for applications where semi-insulating behavior is desired.

**Interpretable analysis of self-driving sputtering with random forest regression**

To move beyond black-box optimization and identify which growth parameters dominate the $E_U$ landscape and how they interact, we next analyze the growth-parameter space explored by the self-driving campaign using an interpretable surrogate model. Specifically, we constructed a random-forest surrogate[70-73] that reproduces the measured $E_U$ landscape over the BO-explored parameter region and performed a suite of analyses to extract salient structure and distill the model into a compact, interpretable representation (see Methods section "Interpretable surrogate decomposition using random forests"). The random forest surrogate trained on all 66 growth runs (including the optimum at 56th run) effectively captures the observed $E_U$ variations across the four-dimensional growth-parameter space, achieving an in-sample coefficient of determination $R^2 = 0.92$ with a root-mean-square error (RMSE) of 36 meV (Fig. 5a), whereas linear and quadratic polynomial regressions yield much poorer performance [$R^2 = 0.11$, RMSE = 121 meV for the linear model (Fig. 5b); $R^2 = 0.41$, RMSE = 98.9 meV for the quadratic model (Fig. 5c)]. Taken together, these comparisons show that classical linear or quadratic regression is insufficient to capture the complex dependence of $E_U$ on the sputter-growth parameters, whereas the BO-discovered behavior can be summarized for interpretation by a compact, data-driven surrogate that we can interrogate to reveal the underlying growth mechanisms.

We next used the surrogate model to identify which process parameters most strongly govern $E_U$ at the dataset level, using feature-importance scores, which capture how much each input contributes to reducing the model's prediction error[70,72]. The resulting importances indicate that the substrate temperature $T$ is the most influential knob (= 0.39), followed by $P_{RF}$ (= 0.21), $F_{Ar}$ (= 0.21), and $F_{O_2}$ (= 0.20) (Fig. 5d). Thus, temperature acts as the primary control parameter, while RF power and the gas flows serve as secondary tuning knobs of comparable strength. To visualize how each parameter influences $E_U$ more intuitively, we computed one-dimensional (1D) partial dependence plots (PDPs) of the random forest model (Figs. 5e-5h). A PDP provides a regime-averaged $E_U$-parameter response curve predicted by the surrogate model, obtained by sweeping one parameter across its range while averaging over the remaining high-dimensional parameter space[70,74] (see Methods section "Interpretable surrogate decomposition using random forests"). These PDPs reveal that $E_U$ exhibits a pronounced minimum in an intermediate range for each knob: intermediate $T$ around 360–540°C, $P_{RF}$ near 100–125 W, $F_{Ar}$ around 15–30 sccm, and a small but finite $F_{O_2}$ of 1–2 sccm. In contrast to $F_{O_2}$, which requires fine tuning at the 1 sccm level, the PDP for $F_{Ar}$ shows a broad low-$E_U$ plateau spanning ~15–30 sccm. This indicates that precise adjustment of the Ar flow is not critical as long as extreme under- or over-supply is avoided. Moving too far away from these intermediate values in any direction leads to an increase in $E_U$. In this sense, the random forest analysis provides a compact, human-interpretable summary of the trends learned implicitly by BO during the closed-loop optimization.

To further examine how the primary and secondary knobs interact, we computed two-dimensional (2D) PDPs as functions of substrate temperature and one additional parameter ($T$-$P_{RF}$, $T$-$F_{Ar}$, and $T$-$F_{O2}$) (Figs. 6a-6c). In all cases, temperature sets the overall concave shape of the $E_U$ landscape, producing a broad minimum at intermediate $T$ (360–540°C). Varying RF power or gas flows mainly shifts $E_U$ up and down, without changing the shape or location of this temperature minimum, indicating weak interactions and an approximately additive response. We quantify this near-additive behavior by constructing an explicitly additive approximation of the random forest surrogate, where the predicted $E_U$ is expressed as the sum of four 1D PDPs;

$$E_U^{add}(T, P_{RF}, F_{Ar}, F_{O_2}) \approx g_T(T) + g_P(P_{RF}) + g_{Ar}(F_{Ar}) + g_{O_2}(F_{O_2}) + c_0, \quad (1)$$

where $g_T(T)$, $g_P(P_{RF})$, $g_{Ar}(F_{Ar})$, and $g_{O_2}(F_{O_2})$ are the 1D PDPs of the random forest surrogate with respect to each growth parameter. The constant $c_0$ was chosen to minimize the mean-squared error of Eq. (1) against the full random-forest predictions over the $N = 66$ BO-acquired samples. This additive PDP approximation [Eq. (1)] preserves much of the structure learned by the full random forest model, reproducing its outputs with $R^2 = 0.84$ (and explaining $R^2 = 0.69$ of the variance in the measured $E_U$). These results indicate that most of the four-dimensional $E_U$ landscape is governed by approximately independent, additive contributions of the four knobs ($T$, $P_{RF}$, $F_{Ar}$, and $F_{O_2}$), while the remaining discrepancy reflects non-additive interactions and/or higher-order structure.

To quantify the remaining non-additive component beyond this largely additive structure, we computed pairwise interaction strengths using Friedman's $H$-statistic[75] from the random forest surrogate (see Methods section "Interpretable surrogate decomposition using random forests"). For a given pair of parameters, $H^2$ quantifies the fraction of the 2D-PDPs' variation that cannot be explained by the sum of the two corresponding 1D PDPs; $H^2 = 0$ corresponds to a purely additive dependence with no interaction, whereas $H^2 = 1$ indicates that non-additive interactions dominate. In our case, all interaction fractions $H^2$ were found to be below 0.12 (Fig. 6d), confirming that interaction effects are generally weak across most parameter pairs. The strongest interactions occur for the $T$-$F_{O_2}$ pair ($H^2 = 0.12$), while pairs involving RF power and Ar gas flow show weaker interactions ($H^2 = 0.017$–$0.043$). Taken together with the additive approximation in Eq. (1), these results motivate a compact and interpretable representation of the random forest surrogate in which the response is predominantly additive, with the dominant residual non-additivity confined to the $T$-$F_{O_2}$ pair:

$$E_U^{total}(T, P_{RF}, F_{Ar}, F_{O_2}) = g_T(T) + g_P(P_{RF}) + g_{Ar}(F_{Ar}) + g_{O_2}(F_{O_2}) \\ + g_{T,O_2}^{int}(T, F_{O_2}) + c_1, \quad (2)$$

where $c_1$ is an offset that absorbs the arbitrary baselines. The interaction term $g_{T,O_2}^{int}(T, F_{O_2})$ is defined as the pure two-dimensional interaction component extracted from the 2D PDP as follows:

$$g_{T,O_2}^{int}(T, F_{O_2}) = g_{T,O_2}(T, F_{O_2}) - g_{O_2}(F_{O_2}) - g_T(T) + c_{T,F_{O_2}}, \quad (3)$$

where $g_{T,O_2}(T, F_{O_2})$ denotes the 2D PDP of the random forest surrogate with respect to $T$ and $F_{O_2}$, and $c_{T,F_{O_2}}$ is the constant chosen such that $\langle g_{T,O_2}^{int}(T, F_{O_2}) \rangle = 0$ over the 2D PDP grid. Incorporating this single interaction term increases the fidelity to the full random forest predictions, evaluated at the $N = 66$ BO-acquired sample points, to $R^2 = 0.90$ and yields $R^2 = 0.75$ against the measured $E_U$, highlighting that the $E_U$ landscape admits an interpretable form: substrate temperature $T$ acts as the dominant knob setting the overall concave shape, while the remaining parameters are largely additive and weakly coupled ($H^2 = 0.017$–$0.043$). The only appreciable deviation from additivity is a modest $T$-$F_{O_2}$ interaction term ($H^2 = 0.12$), plausibly attributable to coupled thermodynamic/kinetic effects (e.g., changes in oxygen chemical potential, surface stoichiometry, and defect incorporation)[76,77] that are not captured by one-dimensional trends alone.

Figure 6e visualizes the role of the $T$-$F_{O_2}$ interaction around the optimum by plotting $g_{T,O_2}^{int}(T, F_{O_2})$ in the restricted window. The interaction term reduces $E_U$ along the same $T$-$F_{O_2}$ ridge where $E_U$ attains its minimum. Thus, in addition to the nearly additive 1D contributions, a modest $T$-$F_{O_2}$ interaction further reduces $E_U$ by approximately 40 meV near the optimal growth conditions, reflecting a complex thermodynamic behavior that is difficult to predict quantitatively from classical physical theory[76,77].

These insights suggest a simple, human-executable optimization strategy for sputter epitaxy of $\beta$-Ga$_2$O$_3$ that mirrors the behavior learned by the self-driving system. Starting from an arbitrary but reasonable set of growth conditions, the largely additive structure of the growth-parameter space indicates that optimization can proceed efficiently through sequential one-dimensional tuning of individual parameters. In practice, this corresponds to first adjusting the substrate temperature to minimize $E_U$, followed by independently tuning secondary parameters such as RF power and the Ar and O$_2$ flow rates, each of which exhibits a largely independent, well-behaved optimum. Combining the individually optimized parameter values already places the process close to the global optimum in $E_U$. From this neighborhood, a more focused two-dimensional optimization in the $T$-$F_{O_2}$ plane is sufficient to capture the residual non-additive contribution that may arise from the thermodynamic stability and/or surface kinetics. In this way, the nominally black-box BO optimization, together with the random forest-based analysis, is distilled into an interpretable and practically actionable optimization strategy that can be implemented by human experimenters without requiring an autonomous platform. Overall, these results demonstrate that the self-driving sputtering framework not only identifies high-quality growth conditions, but also generates a dataset rich enough to enable subsequent interpretable analysis of the inferred process landscape.

**Discussion**

In this work, we demonstrate that RF magnetron sputtering, when combined with an interpretable self-driving growth framework and automated optical analyses, can produce single-phase heteroepitaxial $β$-Ga$_2$O$_3$ films with high crystallographic and optical quality approaching that typically associated with CVD- and MBE-based epitaxy. Using the $E_U$ as a quantitative objective, the closed-loop campaign achieved $E_U$ = 182 meV, the lowest value reported for sputtered $β$-Ga$_2$O$_3$ films and below a representative MOCVD-grown report (220 meV)[46] (Table 1). While this performance remains above the best reported epitaxial films (~150 meV)[48,59] and bulk single crystals (60–140 meV)[48], the residual gap provides a clear roadmap for further improvement, namely, suppressing sub-gap disorder/defect states that are still detectable optically even in single-crystal films. Importantly, the same BO-identified growth window discovered on C-plane Al$_2$O$_3$ is directly transferable to $β$-Ga$_2$O$_3$ substrates, enabling single-crystal homoepitaxy. Given that RF sputtering is a mature, low-operating-cost technology already used for large-area oxide deposition in semiconductor and display manufacturing, these results point to self-driving sputter epitaxy as a promising, industry-compatible route for ultra-wide-bandgap $β$-Ga$_2$O$_3$ epitaxial layers and devices.

Although the initial dataset in our campaign happened to include a sample with a moderately low $E_U$ (< 350 meV), Bayesian optimization is not predicated on starting from a near-optimal recipe. Instead, the exploration-exploitation trade-off encourages broad sampling early on while progressively refining around promising regions. Consistent with this behavior, the $E_U$ values span a wide range across runs while the best-achieved $E_U$ decreases steadily. In our self-driving sputtering system, we further improved robustness against local trapping by adopting an adaptive prior-mean strategy, in which the GP prior mean is resampled at each iteration from a uniform distribution bounded by the minimum and maximum observed $E_U$ values[45]. This design promotes exploration of previously unseen regions and helps the search escape locally optimal regimes. The effectiveness of this prior-mean randomization, both for encouraging exploration and for reproducibly reaching a global optimum regardless of the choice of initial samples, has been validated on simulated data[45].

A distinctive aspect of this study is that self-driving optimization is not treated as a purely black-box procedure. While BO efficiently discovers growth conditions that minimize $E_U$, process development ultimately requires a physical insight and an interpretable understanding of which parameters matter and how they should be adjusted. To bridge autonomous search and human-intuitive tuning, we analyzed the growth-parameter landscape explored by BO and distilled the closed-loop dataset into actionable process knowledge. Using a random forest as an interpretable local surrogate, we identified a near-additive $E_U$ landscape with modest, physically plausible interactions dominated by a temperature and $O_2$ flow coupling. This structure yields a practical tuning logic consisting of sequential one-dimensional adjustments followed by a focused two-dimensional refinement in temperature and $O_2$ flow, and it is consistent with the experimentally observed cross-substrate transferability from heteroepitaxy to single-crystal homoepitaxy.

Importantly, these conclusions are not specific to the choice of surrogate model. For acquisition-driven exploration, we used a GP surrogate to leverage sample efficiency and predictive uncertainty, whereas for interpretable surrogate decomposition, we used a random forest because of its robustness to non-smooth responses and experimental noise, together with its mature interpretability tools. Here, the random forest is used not as a universal predictive model but as an interpretable local surrogate, aimed at exposing dominant trends and modest interactions supported by the BO trajectory rather than out-of-window generalization. Consistent with the random forest analysis, fitting a GP with the prior mean set to the data mean produced the same qualitative picture: additive trends alone were insufficient, whereas including a single temperature and $O_2$ flow interaction term substantially improved the approximation (Supplementary Information, Sec. IV). Looking forward, SDLs can further leverage their high experimental throughput to complement BO with space-filling sampling, enabling surrogate models with higher predictive fidelity across the full parameter space and more robust, transferable interpretable growth rules. Together, these results outline a general path toward interpretable SDLs, in which autonomous experimentation and surrogate-based distillation transform closed-loop optimization from a black-box procedure into physically meaningful, human-usable, and transferable process knowledge and growth rules.

**METHODS**

**Self-driving sputtering system**

$β$-$Ga_2O_3$ thin films were deposited using a custom-ordered sputtering platform (ULVAC, Inc., Japan) equipped with a $Ga_2O_3$ sputtering target. 2-inch C-plane $Al_2O_3$ wafers (K&R Co., Ltd.) were used as received for heteroepitaxy, and Fe-doped semi-insulating $β$-$Ga_2O_3$ substrates (Novel Crystal Technology, Inc.) were used as received for homoepitaxy. The system was operated in a closed-loop, self-driving mode, where deposition parameters were programmatically updated between consecutive runs based on an optimization algorithm (see Methods section "Bayesian optimization with experimental failures and adaptive GP prior mean"). The sputtering chamber was evacuated to a base pressure below $2\times10^{-4}$ Pa before deposition, and films were grown in an $Ar/O_2$ mixture, whose flow rates were controlled by mass flow controllers. The substrate temperature was calibrated by pyrometry, and the thermocouple reading was used for temperature monitoring during growth. The target-to-substrate distance was 100 mm. The deposition time was determined from the pre-calibrated deposition rate to obtain a nominal film thickness of 55 nm. The platform is equipped with a load-lock chamber capable of holding up to five 2-inch wafers. Substrate transfer between the load-lock and the deposition chamber, including transfer after film growth back to the load-lock, was automatically performed using a robotic arm.

Ex situ optical characterization was performed using a UV-Vis-NIR spectrophotometer (V-770; JASCO Corporation, Japan). After each deposition run, the sample was manually transferred from the load-lock to the optical measurement stage (in air). Once optical measurement was initiated, subsequent steps, including automated extraction of the Urbach energy ($E_U$) from the transmittance spectra, programmatic updating of the deposition parameters, and initiation of the next growth run, were executed in an automated manner (see Methods section "Automated optical analysis"). In the present configuration, manual handling is required only for transferring the sample from the load-lock to the optical measurement stage; automating this step would enable fully autonomous operation for up to five consecutive runs per load-lock loading.

**Automated optical analysis**

Optical bandgap energy ($E_g$) and $E_U$ were automatically extracted from the measured optical transmittance spectra $Tr(E)$ using a custom analysis script. The extracted $E_U$ was then forwarded to an automated dataset update routine used for subsequent Bayesian optimization (BO). The absorption coefficient $\alpha(E)$ was estimated from the Beer–Lambert relation[78]. In the simplest approximation (neglecting reflectance and interference fringes),

$$\alpha(E) = -\frac{1}{d} ln[Tr(E) + \varepsilon], \quad (4)$$

where $d$ is the film thickness and $\varepsilon$ is a small constant to avoid singularities at $Tr \to 0$. $E_g$ was then obtained from a Tauc plot assuming a direct bandgap structure[79],

$$(\alpha h\nu)^2 \propto h\nu - E_g, \quad (5)$$

by linear fitting in the appropriate region and taking the intercept with the $h\nu$ axis (see Supplementary Fig. 6a). Although $\beta$-Ga$_2$O$_3$ is commonly regarded as an indirect-bandgap semiconductor, first-principles calculations show that the minimum direct bandgap is only tens of meV above the indirect fundamental bandgap[80], which explains the strong near-edge absorption. Consistent with this, Ricci et al. pointed out that the near-edge optical behavior is effectively that of a Γ direct-like[81], and direct-allowed Tauc analyses are therefore widely used in practice for $\beta$-Ga$_2$O$_3$ thin films[42]. $E_U$ was extracted from the exponential absorption Urbach tail[43] just below the estimated $E_g$ by fitting

$$ln[\alpha(E)] = \frac{1}{E_U} E + C, \quad (6)$$

where $C$ is a constant (see Supplementary Fig. 6b). Details of the automated fitting procedure (including window selection criteria and parameter settings) are available via the Code availability section.

**Bayesian optimization with experimental failures and adaptive GP prior mean**

The BO objective was to minimize the $E_U$. To address practical stagnation in self-driving sputtering, namely, missing/NaN evaluations caused by growth failures under off-optimum conditions and trapping in locally optimal regions, we developed a Gaussian-process (GP)-based BO loop featuring two key innovations: NaN handling by worst-value imputation[44] and adaptive modification of the GP prior mean[45]. After each growth, the measured transmittance spectrum was analyzed to obtain $E_U$, and the resulting value was fed back to a BO algorithm to determine the next set of growth parameters. The parameter vector was defined as $\boldsymbol{x} = (T\ [°C], P_{RF}\ [W], F_{Ar}\ [sccm], F_{O_2}\ [sccm])$, with bounds $T \in [30, 700]$, $P_{RF} \in [50, 150]$, $F_{Ar} \in [10, 50]$, and $F_{O_2} \in [0, 10]$. At iteration $n$, a GP surrogate model was fitted to the dataset $D_{n-1} = \{(\boldsymbol{x}_i, \boldsymbol{y}_i)\}_{i=1}^{n-1}$, where $y_i$ denotes the observed $E_U$, i.e., $y_i = E_U(\boldsymbol{x}_i)$, obtained from the previous iterations of the self-driving loop. Although BO is commonly formulated as a maximization problem, minimization was handled by negating the observations when evaluating the acquisition function.

To improve robustness in closed-loop experiments, we implemented the two mechanisms described above. First, when a sputtering growth failed and $E_U$ was unavailable (NaN) due to growth parameters that are far from optimal, the trial was not discarded; instead, we imputed the objective by assigning the worst (largest) experimental $E_U$ value at that time[44]:

$$\tilde{y}_{n'} = \begin{cases} y_n & \text{if } y_n \neq \text{NaN}, \\ \max_{1 \leq i < n} \tilde{y}_i & \text{if } y_n = \text{NaN}. \end{cases} \quad (7)$$

This imputation of the dataset $D_n = \{(\boldsymbol{x}_i, \boldsymbol{y}_i)\}_{i=1}^{n}$ by $\widetilde{D}_n = \{(\boldsymbol{x}_i, \tilde{\boldsymbol{y}}_i)\}_{i=1}^{n}$ enables direct exploration of a wide parameter space while avoiding experimental failure. Second, we used an adaptive prior mean strategy for the GP mean function[45]. In the GP, the predictive mean depends on the prior mean function $\eta(\boldsymbol{x})$ (here taken as a constant $\eta(\boldsymbol{x}) = m_0$). The prior mean hyperparameter was resampled at each iteration from a uniform distribution bounded by the minimum and maximum of the observed values:

$$m_0^n = \mathcal{U}\left(\min_{1 \leq i < n} \tilde{y}_i, \max_{1 \leq i < n} \tilde{y}_i\right), \quad (8)$$

which helps the search escape locally optimal regions and promotes exploration of previously unseen regions.

The next growth condition $\boldsymbol{x}_{n'+1}$ was chosen by maximizing the expected improvement (EI) acquisition function computed from the GP predictive distribution. As a practical GP configuration, we adopted a Matérn 5/2 kernel (commonly used for functions with relatively steep gradients)[82] and included a constant observation-noise variance term to account for run-to-run variability.

**Interpretable surrogate decomposition using random forests**

Random forest regression was implemented using scikit-learn's RandomForestRegressor[83] and trained on $N = 66$ samples acquired by the self-driving sputtering. Hyperparameters were tuned using 5-fold cross-validation on the BO-acquired dataset. The final model used for interpretability employed 300 trees with unlimited depth, a minimum of one sample per leaf, and the square root of the number of available features considered at each split. This model achieved an in-sample $R^2 = 0.92$, while a 5-fold cross-validation yielded a more conservative $R^2 \approx 0.43$, reflecting the limited number and BO-biased distribution of samples in parameter space. Accordingly, the random forest is used in this work primarily as an interpretable surrogate to summarize the $E_U$ landscape within the explored growth window, rather than as a high-accuracy predictive model for arbitrary unseen conditions.

Let $\hat{f}(T, P_{RF}, F_{Ar}, F_{O_2})$ denote the trained random forest regression function. Feature importance and partial dependence analyses were performed on $\hat{f}$. Feature importances were obtained from the impurity-based importance scores, where the decrease in mean squared error at each split is accumulated for the corresponding parameter and averaged over all trees, followed by normalization so that the importances of all features[70,71].

1D partial dependence functions $g_s(z_s)$ were defined for each input parameter $s \in \{T, P_{RF}, F_{Ar}, F_{O_2}\}$ and parameter value $z_s$ as

$$g_s(z_s) = \frac{1}{N}\sum_{i=1}^{N} \hat{f}(z_s, x_{i,-s}), \tag{9}$$

where $x_{i,-s}$ denotes all parameter values of sample $i$ (from 1 to $N = 66$) except $x_{i,s}$. Here, $x_{i,s}$ is the value of parameter $s$ value in sample $i$. Similarly, for a pair of parameters $(s, t)$ and parameter values $z_s$ and $z_t$, the 2D partial dependence function was defined as

$$g_{s,t}(z_s, z_t) = \frac{1}{N}\sum_{i=1}^{N} \hat{f}(z_s, z_t, x_{i,-s,-t}), \tag{10}$$

where $t \in \{T, P_{RF}, F_{Ar}, F_{O_2}\}$ and $x_{i,-s,-t}$ denotes all parameter values of sample $i$ except $x_{i,s}$ and $x_{i,t}$. Following Eq. (9), each 1D PDP was evaluated on a grid spanning the observed range of $s$. 2D PDPs were also calculated in the same manner for selected pairs of parameters, such as $(T, P_{RF})$.

Pairwise interaction strength between input parameters was quantified using Friedman's $H^2$ interaction statistic[75] based on the above partial dependence functions. For a given parameter pair $(s, t)$, the interaction component was defined as

$$g_{s,t}^{int}(z_s, z_t) = g_{s,t}(z_s, z_t) - g_s(z_s) - g_t(z_t) + c_{s,t}, \tag{11}$$

where $c_{s,t}$ is the constant chosen such that $\langle g_{s,t}^{int}(z_s, z_t) \rangle = 0$ over the 2D PDP grid. Friedman's $H_2$ statistic for the parameter pair $(s, t)$ was then calculated as

$$H_{s,t}^2 = \frac{\sum g_{s,t}^{int}(z_s, z_t)^2}{\sum (g_{s,t}(z_s, z_t) - c_{s,t})^2}, \tag{12}$$

where the sums over $(z_s, z_t)$ approximate the corresponding integrals using the same parameter grid as that employed for computing the 2D PDPs.

**Author contributions**

Y.K.W. conceived the idea and directed and supervised the project. Y.K.W. and Y.O. designed the interpretable self-driving sputtering system. Y.O. established the sputtering and optical measurement setup. Y.K.W. and T.O. implemented the Bayesian optimization algorithm. Y.K.W. implemented the automated optical analysis algorithm. F.B.R. operated the self-driving sputtering system. F.B.R. and Y.K.W. carried out the additional sample characterizations. Y.K.W. performed the interpretable random forest-based analysis. Y.K.W., Y.O., F.B.R., T.O., and Y.T. discussed the results. Y.K.W. wrote the manuscript with input from all authors.

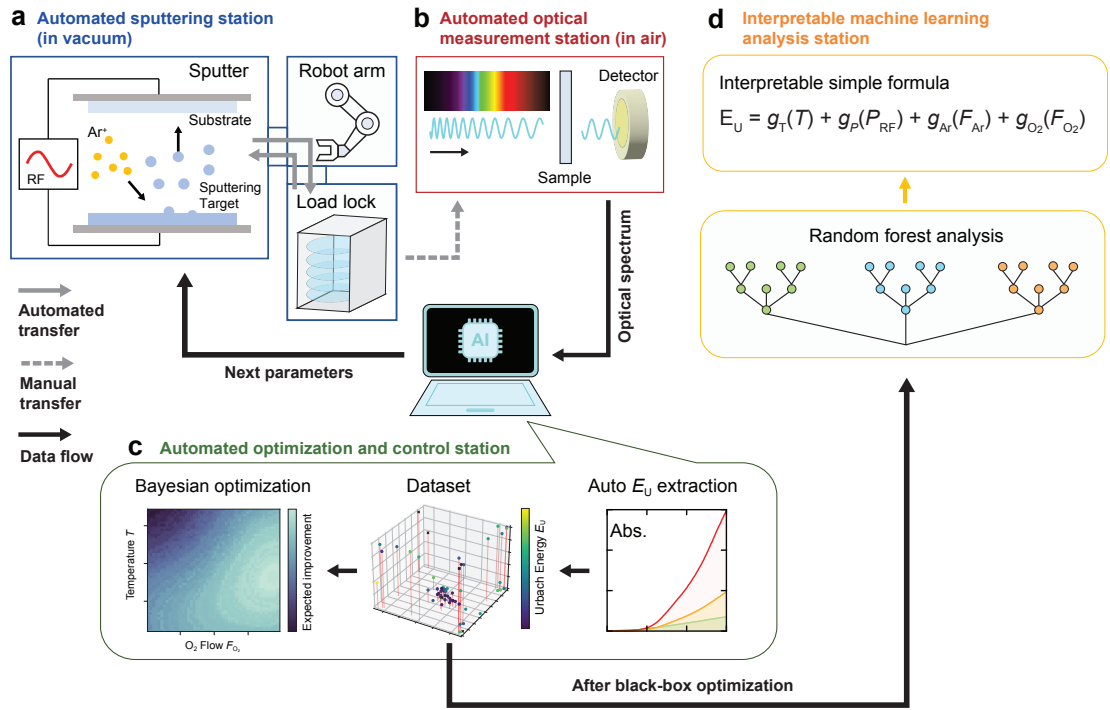

**Fig. 1. Closed-loop self-driving sputter epitaxy with interpretable machine learning.**
**a** Automated RF sputtering station operated in vacuum for thin-film growth. **b** Automated optical measurement station operated in air to acquire optical transmittance spectra. **c** Automated optimization and control station that performs (1) auto-extraction of the Urbach energy $E_U$ from the optical spectra, (2) dataset update, and (3) Bayesian optimization to propose the next growth conditions and automatically triggers the subsequent sputter run. **d** Interpretable ML analysis station, where a random forest model is trained on the accumulated dataset and distilled into human-interpretable relations for process understanding.

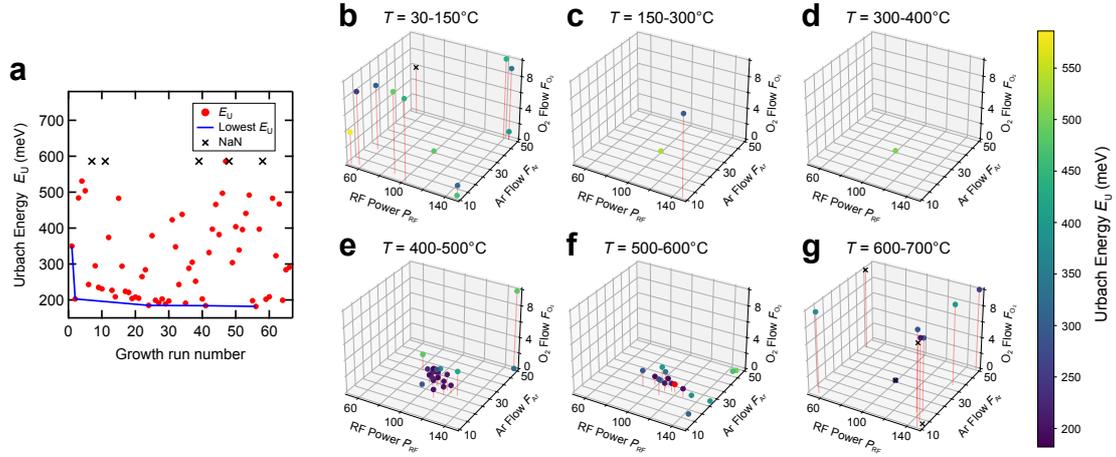

**Fig. 2. Bayesian optimization of $E_U$ over a four-growth parameter space. a** The experimental $E_U$ (red circles) as a function of growth run number. The blue line indicates the lowest $E_U$ achieved up to each run. Black crosses indicate runs for which the $E_U$ could not be extracted (NaN). **b–g**, The experimental $E_U$ in the three-dimensional parameter space $P_{RF}$-$F_{Ar}$-$F_{O_2}$ for different $T$ windows of **b** 30-150°C, **c** 150-300°C, **d** 300-400°C, **e** 400-500°C, **f** 500-600°C, and **g** 600-700°C. Marker color represents $E_U$ (color bar, meV), visualizing how Bayesian optimization concentrates sampling. In **f**, the red circle represents the lowest $E_U$ (182 meV).

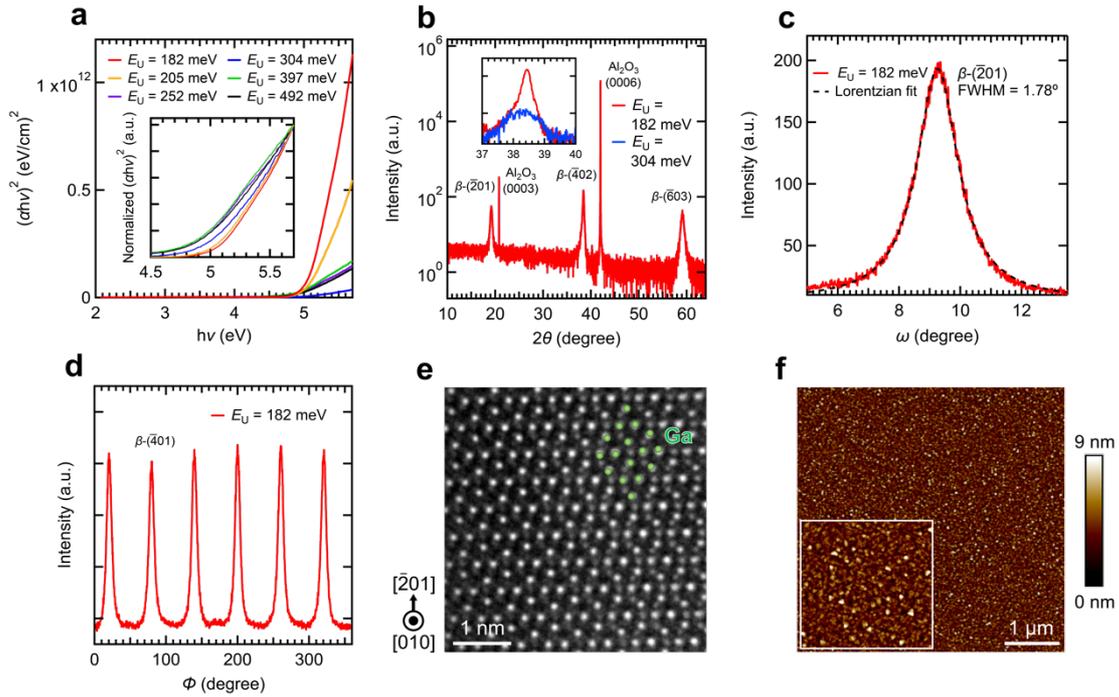

**Fig. 3. Optical and structural properties of the optimized heteroepitaxial *β*-Ga$_2$O$_3$ films on C-plane Al$_2$O$_3$. a** Tauc plots $(\alpha h\nu)^2$ derived from optical transmittance spectra for representative films with different $E_U$ values. The inset in **a** shows normalized $(\alpha h\nu)^2$ near the absorption edge. **b** XRD $\theta$-$2\theta$ scan for the optimized film ($E_U$ = 182 meV). The inset in **b** also shows the scan for the film with $E_U$ = 304 meV. **c** Rocking curve ($\omega$-scan) of the *β*-Ga$_2$O$_3$ ($\bar{2}$01) reflection for the optimized film, with a Lorentzian fit (dashed line). **d** In-plane $\phi$-scan of the *β*-($\bar{4}$01) reflection for the optimized film. **e** Atomic-resolution HAADF-STEM image of the optimized film viewed along the [010] direction of the *β*-Ga$_2$O$_3$ (parallel to the [1$\bar{1}$00] direction of the C-plane Al$_2$O$_3$ substrate). In **e**, green spheres indicate Ga-occupied columns. **f** AFM image (5 μm × 5 μm) of the optimized film. The inset in **f** shows a magnified view (1 μm × 1 μm).

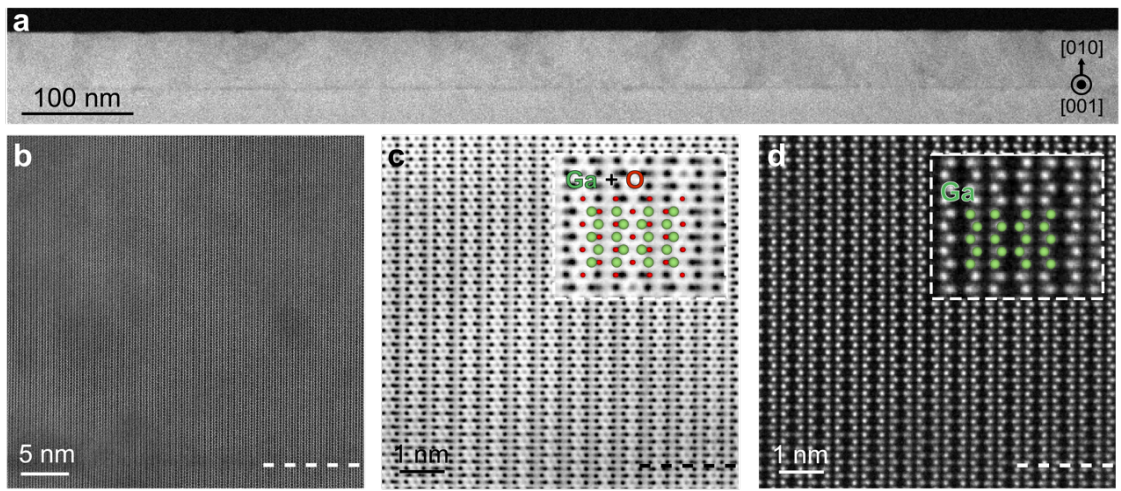

**Fig. 4. Homoepitaxial β-Ga₂O₃ grown under the optimized condition. a** HAADF-STEM image of the homoepitaxial $\beta$-Ga$_2$O$_3$ film viewed along the [001] direction. **b** Magnified HAADF-STEM image near the interface in **a**. **c,d** Magnified **c** ABF- and **d** HAADF-STEM image near the interface in **b**. In **b-d**, dashed lines indicate the interface. The insets in **c** and **d** represent an atomic-resolution magnified image with Ga-occupied (green spheres) and O-occupied (red spheres) columns.

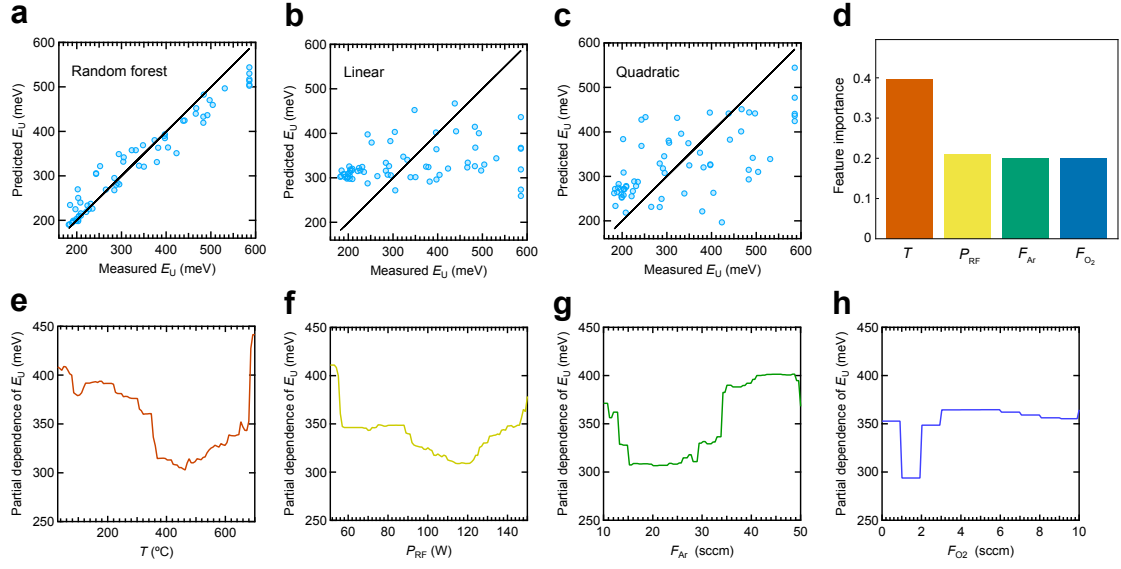

**Fig. 5. Interpretable random forest analysis for single-parameter effects on $E_U$. a-c** Predicted versus measured $E_U$ for three regression models trained on the experimental 66 sample dataset: **a** random forest, **b** linear, and **c** quadratic. The diagonal line indicates perfect agreement. **d** Feature importance obtained from the random forest model, comparing the relative contributions of $T$, $P_{RF}$, $F_{Ar}$, and $F_{O_2}$. **e-h** 1D PDPs from the random-forest model showing the marginal effect of **e** $T$, **f** $P_{RF}$, **g** $F_{Ar}$, and **h** $F_{O_2}$ on $E_U$.

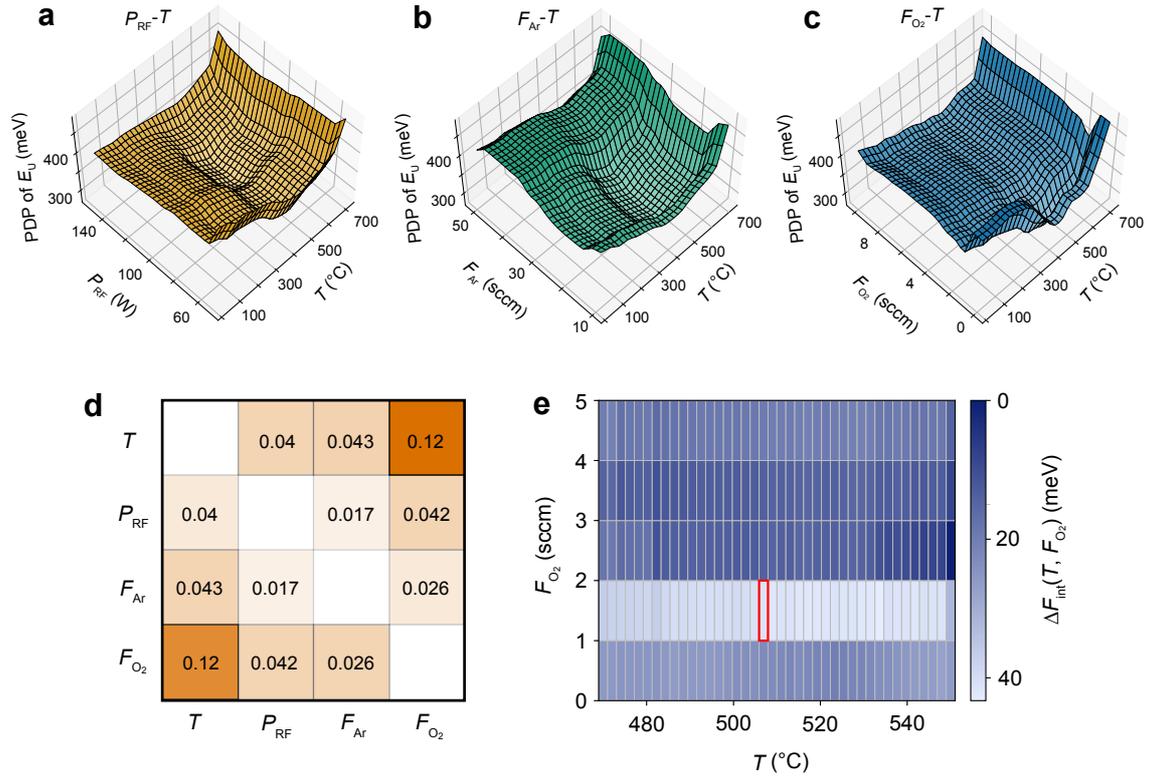

**Fig. 6. Visualizing and quantifying pairwise interactions in the $E_U$ landscape.** **a-c** 2D PDPs of $E_U$ as a function of $T$ and each other parameter: **a** $P_{RF}$, **b** $F_{Ar}$, and **c** $F_{O_2}$. **d** Friedman's $H$-statistic pairwise matrix, quantifying the magnitude of non-additive effects between parameters. **e** $g_{T,O_2}^{int}(T, F_{O_2})$ in the restricted window around the optimum ($T$ = 470–550°C, $F_{O_2}$ = 0–5 sccm). In **e**, the red rectangle marks the region corresponding to the optimized condition.

| Method | $E_U$ (meV) | ($\bar{2}$01) RC-FWHM (°) | AFM RMS (nm) | Ref. |
|---|---|---|---|---|
| RF sputter (this work) | 182 | 1.78 | 1.15–1.27 | This work |
| RF sputter (literature) | 280 | N/A | N/A | 47 |
| Low-pressure (LP) CVD | 150 | 1.18–1.50 | 1.82–3.50 | 48-51 |
| MOCVD | 220–345 | 0.60–2.25 | 1.28–1.69 | 46,52-58 |
| PAMBE | 150 | 1.0 | 0.21–7 | 59-62 |
| MBE | 320 | 0.9–1.9 | 0.88 | 31,63,64 |
| halide-VPE | N/A | 0.33 | N/A | 65 |

**Table 1. Benchmark of optical, structural, and surface quality for *β*-Ga$_2$O$_3$ films on C-plane Al$_2$O$_3$ grown by representative growth methods.** $E_U$, *β*-Ga$_2$O$_3$ ($\bar{2}$01) rocking-curve FWHM, and RMS roughness are compared between this work (RF sputtering) and representative literature reports. Ranges indicate reported values across references; N/A denotes not reported. A more comprehensive list and details are provided in Table S1.